\documentclass[aps,prl,twocolumn,showpacs]{revtex4}
\usepackage{graphicx}% Include figure files
\usepackage{dcolumn}% Align table columns on decimal point
\usepackage{bm}% bold math
\begin{document}
%\draft
\title{Critical point and the nature of the pseudogap of single-layered copper oxide Bi$_{2}$Sr$_{2-x}$La$_{x}$CuO$_{6+\delta}$ superconductors  }

\author{Guo-qing~Zheng$^1$, P. L. Kuhns$^2$, A. P. Reyes$^2$, B. Liang$^3$ and C. T. Lin$^3$}
\affiliation{$^1$Department of Physics, Okayama University,  Okayama 700-8530,
Japan}

\affiliation{$^2$ National High Magnetic Field Laboratory, Tallahassee, FL, USA}
\affiliation{$^3$Max-Planck-Institut fur Festkorperforschung, Heisenbergstr.1, D-70569 Stuttgart, Germany}

\date{Phys. Rev. Lett. 94, 047006 (2005)}
%\twocolumn[
%\widetext
%
\begin{abstract}
We apply strong magnetic fields of $H$=28.5$\sim$43 T to suppress superconductivity (SC) in  the cuprates Bi$_{2}$Sr$_{2-x}$La$_x$CuO$_{6+\delta}$ ($x$=0.65, 0.40, 0.25, 0.15 and 0), and investigate the low temperature ($T$) normal state  by  $^{63}$Cu nuclear spin-lattice relaxation rate ($1/T_1$) measurements. We find that the pseudogap (PG) phase persists deep inside the  overdoped region but terminates at $x \sim$ 0.05 that  corresponds to the hole doping concentration of approximately 0.21. Beyond this critical point, the normal state  is a Fermi liquid characterized by the   $T_1T$=const relation.  
%This is the first phase diagram extending to the zero-$T$ limit for high-$T_c$ superconductors that is obtained by a microscopic probe. 
A comparison of the superconducting state with the $H$-induced normal state in the $x$=0.40 ($T_c$ = 32 K) sample indicates that there remains substantial part of the Fermi surface even in the fully-developed PG state, which suggests that the PG and SC are coexisting matters.
  
\end{abstract}
\vspace*{5mm}
%]
\maketitle
\sloppy
%\narrowtext

In many cases, the normal state of the high transition-temperature ($T_c$) copper oxide (cuprate) superconductors above $T_{c}$  deviates strongly from that described by  Landau's Fermi liquid theory \cite{Anderson}. One of the experimental facts taken as evidence for such deviations is the opening of a pseudogap (PG) above $T_c$, a phenomenon of loss of density of states (DOS) \cite{review}. The pseudogap is pronounced at  low doping level, in the so-called underdoped regime. The pseudogap temperature,  $T^*$, generally decreases as the carrier doping rate increases. However,  it is unclear whether  $T^*$ finally merges into the $T_c$ curve in the overdoped regime \cite{Anderson2}, or it terminates before superconductivity disappears \cite{Orenstein,Tallon}.  Different classes of theories have been put forward to explain the pseudogap phenomenon (for examples, see Ref.\cite{Fukuyama,Lee,Emery,Zhang,Varma,Laughlin,Yanase,Pines}). It is interesting that these theories generally also propose different mechanisms for the occurrence of superconductivity. 
Since the topology of the phase diagram has great impact on the mechanism of the high-$T_c$ superconductivity, it is important to clarify the doping dependence of the pseudogap. Unfortunately, the onset of superconductivity, typically at $\sim$100 K, and the large upper critical field $H_{c2}$ ($\sim$ 100 T) prevents investigation of how the pseudogap evolves with doping. The highest static field available to date ($\sim$ 30 T) was only able to reduce $T_{c}$ to half its 
value at most \cite{Zheng1,Zheng2}. Even the pulsed magnetic field is not enough to suppress superconductivity completely \cite{Boebinger}.

Meanwhile,  from  angle resolved photoemission spectroscopy (ARPES),  it was found that below $T^*$ the Fermi surface is progressively destroyed with lowering the temperature and  there remains only four arcs at the Fermi surface at $T=T_c$ \cite{Norman}.  It would be helpful to see how  these arcs would evolve if the superconductivity is removed. But again the robust superconducting phase makes it difficult to reveal the properties of the low temperature pseudogap state. 
 
Here we address these two issues by using  single layered cuprates, Bi$_{2}$Sr$_{2-x}$La$_{x}$CuO$_{6+\delta}$,  which have  substantially lower $T_c$ and $H_{c2}$. We study the property of the ground state induced by the application of magnetic fields of 28.5$\sim$43 T, by using nuclear magnetic resonance (NMR) technique. This system is suitable for such study for it can be tuned from the overdoped regime to the underdoped regime by replacing La for Sr, and very highly overdoped  by replacing Pb for Bi \cite{Maeda,Wang}. Moreover, it has been long suspected that interlayer coupling could complicate the superconducting-state properties as well as the normal-state properties. The present system helps since it has only one CuO$_2$ plane in the unit cell. 
This material has additional advantage in its nearly ideal two dimensional structure with the largest transport anisotropy (10$^4$-10$^5$) among known cuprates \cite{Martin}.
 We were able to suppress superconductivity completely in the samples of $x$=0.40, 0.25, 0.15 and 0, which are in the optimally doped to overdoped regimes, by applying  magnetic fields of 28.5$\sim$43 T generated by the Bitter and Hybrid  magnets in the National High Magnetic Field Laboratory, Tallahassee, Florida. 
 
% We find that the pseudogap is remarkbly large for $x$=0.65 with $T^*$=200 K, persisting into the overdoped region ($T^*$=60 K for $x$=0.15), but terminates at around $x$=0.05 which approximately corresponds to the hole concentration of $p_{cr}$=0.21.   Comparison of  the low-$T$ normal state where superconductivity is suppressed with the superconducting state indicates that there remains substantial DOS at the Fermi level even in the well-developed pseudogap state, which suggests that superconductivity and the pseudogap are coexisting matters.  Our results establish, for the first time,  the zero-$T$ limit phase diagram for the pseudogap regime.  

Single crystals of Bi$_{2}$Sr$_{2-x}$La$_{x}$CuO$_{6+\delta}$ were grown by the traveling solvent floating zone (TSFZ) method with starting materials of Bi$_2$O$_3$, SrCO$_3$, La$_2$O$_3$ and CuO (Ref. \cite{Lin}). Compositional measurement was performed by  Auger electron spectroscopy with an error of $\pm$2 wt.\%. The excess oxygen $\delta$ resides on the Bi$_2$O$_2$ block and is believed to be responsible for the carrier doping in the CuO$_2$ plane. 
The amount of $\delta$ of the present samples was estimated to be 0.36 as described in detail in Ref. \cite{Lin}. Replacing La for Sr removes holes from the CuO$_2$ plane and increases $T_c$. The $T_c$ of Bi$_{2}$Sr$_{2}$CuO$_{6.36}$ without  La-doping is found to be 8 K.   The maximal $T_c$=32 K was obtained for La concentration of  $x$=0.4, which is in good agreement with that reported in Ref.  \cite{Ono}.

For NMR measurements, two or three single crystal platelets  with the dimensions of  15$\times$5$\times$1 mm$^{3}$ were  aligned along the  $c$-axis.  For all measurements, the external field is applied along the $c$-axis. A standard phase-coherent pulsed NMR spectrometer was used to collect data. The NMR spectra were obtained by sweeping the magnetic field at a fixed frequency (325$\sim$492 MHz) and recording the size of the spin echo area. 

The full width at the half maximum (FWHM) of the $^{63}$Cu NMR line for the central transition ($m=1/2\leftarrow \rightarrow m=-1/2$ transition) at $T$=4.2 K  is 1.8 kOe for $x$=0 but decreases with increasing $x$, reducing to  1.0 kOe for $x$=0.4.  This is probably due to  removal of  modulation in the Bi$_2$O$_2$ block that is commonly seen in Bi-based cuprates \cite{Matsui}. The  $^{63}$Cu nuclear spin-lattice relaxation rate, $1/T_1$, was measured at the spectrum peak by using a single saturation pulse and fitting the recovery of the nuclear magnetization ($M(t)$) after the saturation pulse to the theoretical curve given by Narath \cite{Narath}: $\frac{M(\infty)-M(t)}{M(\infty)}=0.1exp(-t/T_1)+0.9exp(-6t/T_1)$. The fitting is satisfactorily good in the whole temperature range and at all fields.

\begin{figure}
\begin{center}
\includegraphics[scale=0.5]{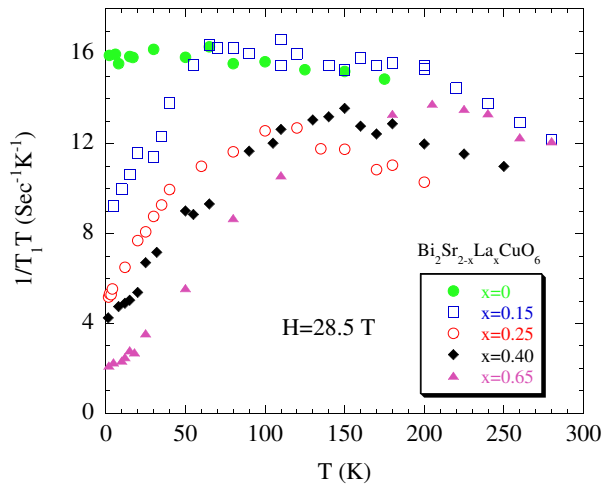}
\caption{(Color on-line) The quantity $1/T_1T$ plotted  against $T$ for Bi$_{2}$Sr$_{2-x}$La$_{x}$CuO$_{6+\delta}$ measured at a field of 28.5 T applied along the c-axis.}
\label{fig:1}
\end{center}
\end{figure}

Figure 1 shows the temperature dependence of $1/T_1T$ for various doping concentrations. Upon reducing $T$ from around room temperature, there is a general trend that $1/T_1T$ increases for all concentrations. For $x$=0, below $T\sim$100K, however, $1/T_1T$=const,  a relation commonly seen in conventional metals. In contrast, for $x\geq$0.15, there appears a broad peak at $T^*$=60$\sim$200 K, depending on $x$.

In general, $1/T_1T$ is related to the dynamical susceptibility 
 $\chi (q, \omega)$ as
\begin{eqnarray}
  1/T_{1}T = \frac{3k_{B}}{4}\frac{1}{\mu_{B}^{2}\hbar^{2}}\sum_{q}A_{q}A_{-q}\frac{\chi^
{"}(q,\omega)}{\omega} 
\end{eqnarray}
where $A_{q}$ is the $q$-dependent
hyperfine coupling constant \cite{Moriya}. In conventional metals, both $A_{q}$ and $\chi (q)$ are basically $q$-independent so that eq. (1) yields to a $T_1T$=const relation.  In most high-$T_c$ cuprates,  the dynamical susceptibility has  a peak at the antiferromagnetic wave vector $Q$=($\pi$,$\pi$). $1/T_1T$ is then shown to be proportional to $\chi_{Q}$. The increase of $1/T_1T$ upon decreasing temperature is generally attributed to the increase of $\chi_{Q}$, namely, to the development of antiferromagnetic correlations.  
For antiferromagnetically correlated metals, this quantity  follows a Curie-Weiss relation  \cite{Millis2,Moriya2}, $\chi_{Q}\propto 1/(T+\theta)$, so that $1/T_1T \sim 1/(T+\theta)$ before superconductivity sets in. In the $x$=0.65 sample, this is true above $T$=200 K, while below this temperature $1/T_1T$ starts to decrease, leaving a broad peak at around $T^*$=200 K. This is a typical pseudogap behavior seen in this NMR quantity \cite{Yasuoka}. Our observation of the pseudogap in this single-layered cuprate system is consistent with 
that made by the ARPES measurement for a $x$=0.35 sample \cite{Harris}. 
Interestingly, the pseudogap persists even in the $x$=0.15 sample which is in the overdoped regime, although with a reduced $T^*$=60 K. {\it Such a low $T^*$ has not so far been  possible to access}, since it is below $T_c$ in most materials.  As noted already, in the $x$=0 sample, however, the pseudogap is no more present. Instead, a $T_1T$=const relation holds below $T$=100 K, which indicates that the normal state is a Fermi liquid. The result that the magnitude of $1/T_1T$ for $x\leq$0.15 is enhanced over that for $x\geq$0.25 is probably due to the increase of the transferred hyperfine coupling constant which has previously been reported in the heavily overdoped regime \cite{Kitaoka}.

\begin{figure}
\begin{center}
\includegraphics[scale=0.5]{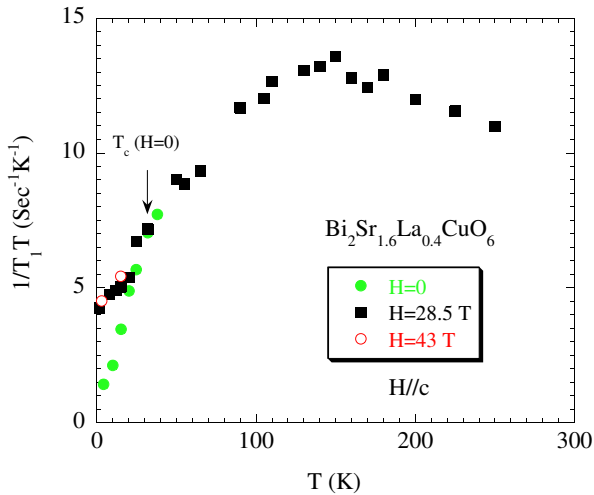}
\caption{(Color on-line) Magnetic field dependence of $1/T_1T$ for  Bi$_{2}$Sr$_{1.6}$La$_{0.4}$CuO$_{6+\delta}$. The arrow indicates $T_c$ at zero magnetic field.}
\label{fig:2}
\end{center}
\end{figure}

Figure 2 shows the magnetic field dependence of $1/T_1T$ for $x$=0.40 under $H$=0, 28.5 T and 43 T. The data for $H$=0 were obtained by NQR (nuclear quadrupole resonance) measurements at the frequency of $\nu_Q\sim$30.2 MHz. The data for $H$=43 T were obtained at the hybrid magnet (outsert field of 11 T and insert field of 32 T) at the High Magnetic Field Laboratory. Note that below $T_c$=32 K, $1/T_1T$ is $H$-dependent between 0 and 28.5 T, but no magnetic field dependence is observed beyond 28.5 T. This indicates that the superconductivity for the $x$=0.40 sample is suppressed by a field greater than 28.5 T, which is also supported by the ac susceptibility measurement using the NMR coil. Therefore, our results for $x\leq$0.40 characterize microscopically the low-$T$ normal (ground) state when superconductivity is removed. 

In Fig. 3 we compare the high field ($H$=28.5 T) data and the zero-field data for the $x$=0 sample. In the normal state above $T_c (H=0)$=8 K, both sets of data agree well. This indicates that the Fermi liquid state in this overdoped sample is an intrinsic property; it is not an effect of high magnetic field. Note that  the Fermi liquid state persists when the superconducting state is suppressed.

\begin{figure}
\begin{center}
\includegraphics[scale=0.5]{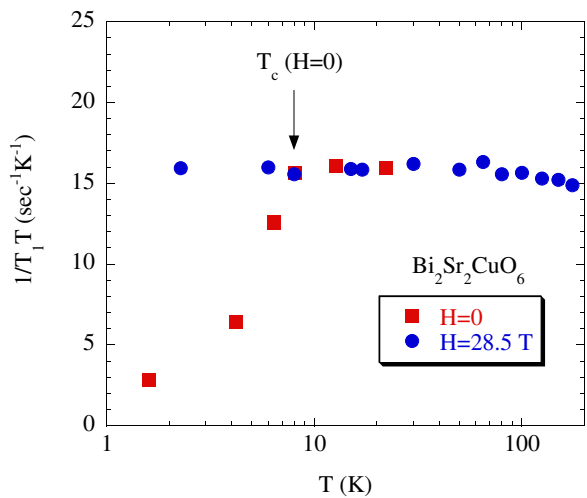}
\caption{(Color on-line) Magnetic field dependence of $1/T_1T$ for the as-grown, overdoped sample, Bi$_{2}$Sr$_{2}$CuO$_{6+\delta}$. The arrow indicates $T_c$ at zero magnetic field.}
\label{fig:3}
\end{center}
\end{figure}
  
The doping dependence of $T^*$ is shown in Fig. 4, along with the $x$-dependence of $T_c$ that was determined as the zero resistance temperature  and agrees well with the onset temperature of the Meissner signal in the ac susceptibility measured using the NMR coil.   The maximal $T_c$ is achieved at $T_c$=32 K for $x_{opt}$=0.40. 
The results  indicate that there exists a critical doping concentration $p_{cr}$ at which the pseudogap terminates and beyond which the ground state when the superconductivity is suppressed is a Fermi liquid.  The critical point is around $x$=0.05 which corresponds to  $p_{cr} \sim$ 0.21, according to Ando's characterization \cite{Ono} (see the upper scale of the transverse axis of Fig. 3). 
We mention a caveat that $T^*$ at zero magnetic field for the overdoped regime  could be slightly higher than that we found here at high magnetic field \cite{Zheng2}, therefore $p_{cr} $ could be slightly higher. However, the limit for largest possible $p_{cr}$ is set by the $x$=0 sample ($p\sim$0.22) which shows no pseudogap.

Note that the $p_{cr}$ we found is much larger than  the optimal doping concentration ($p_{opt} \sim$0.15). Therefore, our results indicate that there is no quantum phase transition taking place at the optimal doping, as opposed to the hypothesis that is frequently conjectured \cite{Orenstein,Sachdev}. However, if the pseudogap is associated with some sort of phase transition \cite{Laughlin}, then $p_{cr} \sim$ 0.21 may be viewed as a quantum critical point. But again, note that $p_{cr}$ is far greater than the optimum doping concentration $p_{opt}=$0.15. It is interesting that many physical quantities, such as the superfluid density  \cite{Panagopoulos}, show distinct change upon crossing a doping concentration that is close to the present $p_{cr}$. 

\begin{figure}
\begin{center}
\includegraphics[scale=0.5]{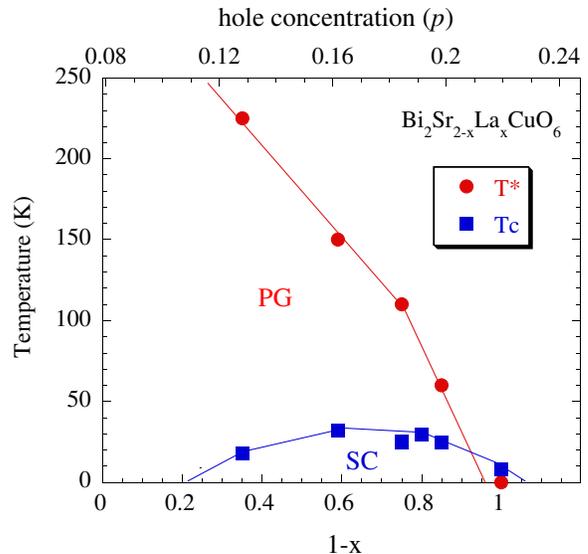}
\caption{(Color on-line) Phase diagram obtained from NMR measurements for  Bi$_{2}$Sr$_{2-x}$La$_{x}$CuO$_{6+\delta}$. $T^*$ is the temperature below which the pseudogap develops, and  $T_{c}$ is the superconducting transition temperature. The upper scale of the transverse axis is adopted from Ref. \cite{Ono}. PG and SC denote the pseudogap phase and superconducting phase, respectively.}
\label{fig:4}
\end{center}
\end{figure}

Finally, the field dependence of $1/T_1T$ below $T_c (H=0)$, as seen in Fig. 2, indicates that the pseudogap is an incomplete gap; even in the fully-developed pseudogap state, {\it i.e.} at $T\sim$1 K, there remains substantial DOS at the Fermi level, which is lost {\it only} after superconductivity sets in. This suggests that superconductivity and pseudogap are coexisting matters. Below $T^{*}$, some parts of the Fermi surface are lost due to the onset of the pseudogap, but other parts of the Fermi surface remain ungapped. If one roughly estimates the DOS from $\sqrt{T_1T^*/T_1T}$, then about 2/3 of the Fermi surface is gapped at $T=T_c$ in the case of Bi$_{2}$Sr$_{1.6}$La$_{0.4}$CuO$_{6+\delta}$, but 1/3 remains ungapped. Only below $T_c$, the remaining Fermi surface is gapped as well due to the onset of superconductivity. Our conclusion that the PG and superconductivity are coexisting matters is also supported by the lack of correlation between $T^*$ and $T_c$.  In the present system, the temperature scale of $T^*$ is the same as that for samples with $T_c$ of $\sim$100 K \cite{review}, even though $T_c$ of the present system is much lower. 
%Therefore,  the pseudogap and superconductivity are coexisting matters. 
The existence of a critical carrier concentration for the pseudogap, the lack of scaling or correlation between $T_c$ and $T^*$, and the persistence of the pseudogap when superconductivity is removed seem not to favor  superconducting precursor as a candidate responsible for the pseudogap which was proposed theoretically by several authors. 

Before closing, we note that coexistence of different states of matter appears to be ubiquitous in various sub-fields of condensed matter physics. For example, our finding of the coexistence of pseudogap and superconducting states bares resemblance to the coexistence of antiferromagnetically-orderd and superconducting states in heavy fermion materials. In the heavy fermion compounds CeRh$_{1-x}$Ir$_x$In$_5$, the sharing (or competing) for the Fermi surface by the two coexisting states also occurs \cite{ZhengPRB}. In fact,  possible similarities between the phase diagram in these two different classes of materials was  pointed out by Laughlin {\it et al} \cite{LaughlinPines}. Therefore, our finding may serve to bridge the understanding of these two sub-fields. 

In summary, we have performed $^{63}$Cu NMR studies at very high  magnetic fields  in the single-layered cuprate superconductors Bi$_{2}$Sr$_{2-x}$La$_{x}$CuO$_{6+\delta}$ for the entire doping regime ($x$=0, 0.15, 0.25, 0.40 and 0.65). The low-$T$ normal state of the  samples with $x\leq$0.40 ($T_c=8\sim$32 K) has been accessed by suppressing superconductivity completely with the strong magnetic fields. It has been found that there exists a critical doping concentration at which the pseudogap state terminates. The critical concentration is $p_{cr}\sim$0.21 ($x\sim$0.05), which is deep inside the overdoped region. Beyond $p_{cr}$, the normal state down to $T$=2 K when the superconductivity is removed is a Fermi liquid state as evidenced by the $T_1T$=const relation. Comparison of  the low-$T$ normal state where superconductivity is suppressed with the superconducting state suggests that superconductivity and the pseudogap coexist. This is also supported by the lack of correlation between $T^*$ and $T_c$.
These results that characterize microscopically, for the first time, the zero-$T$-limit normal state  of the copper-oxide superconductors,  should pose  a constraint on theories for   the high-$T_c$ superconductivity and may also provide clues for understanding other strongly correlated electron systems such as heavy fermion materials.

 We thank Y. Kitaoka for continuous interest and encouragements,  A. Sakai, M. Osada and  M. Kakihana    for  collaboration in the early stage of this work, and  A.V. Balatsky, H. Kohno   and C.M. Varma for stimulating discussions.
This work was supported in part 
 by a Grant-in-Aid for
Scientific Research  from MEXT. NHMFL is supported by NSF and the State of Florida.

\end{document}